\documentclass[manuscript, screen]{acmart}

\usepackage{multirow}
\usepackage{graphicx}
\usepackage{xspace}
\usepackage{tcolorbox}
\usepackage{booktabs}
\usepackage{enumitem}
\usepackage{tikz}
\usetikzlibrary{mindmap,trees}

\AtBeginDocument{%
  }

\setcopyright{acmlicensed}
\acmJournal{TOSEM}

\newcommand{\toolname}{LLM-as-a-Judge\xspace}

\renewcommand\footnotetextcopyrightpermission[1]{}
\begin{document}

\title{From Code to Courtroom: LLMs as the New Software Judges}

\author{Junda He}
\email{jundahe@smu.edu.sg}

\author{Jieke Shi}
\email{jiekeshi@smu.edu.sg}

\affiliation{%
  \institution{Singapore Management University}
  \country{Singapore}
}

\author{Terry Yue Zhuo}\authornote{Corresponding author}
\email{terry.zhuo@monash.edu}

\affiliation{%
  \institution{Monash University \& CSIRO's Data61}
  \country{Australia}
}

\author{Christoph Treude}
\email{ctreude@smu.edu.sg}

\affiliation{%
  \institution{Singapore Management University}
  \country{Singapore}
}

\author{Jiamou Sun}
\email{frank.sun@anu.edu.au}
\affiliation{%
\institution{CSIRO's Data61} 
\country{Australia}
}

\author{Zhenchang Xing}
\email{zhenchang.xing@anu.edu.au}

\affiliation{%
\institution{CSIRO's Data61 \& Australian National University} 
\country{Australia}
}

\author{Xiaoning Du}
\email{xiaoning.du@monash.edu}
\affiliation{%
  \institution{Monash University}
  \country{Australia}
}

\author{David Lo}
\email{davidlo@smu.edu.sg}
\affiliation{%
  \institution{Singapore Management University}
  \country{Singapore}
}

\renewcommand{\shortauthors}{He et al.}

\begin{abstract}
Recently, Large Language Models (LLMs) have been increasingly used to automate SE tasks such as code generation and summarization. However, evaluating the quality of LLM-generated software artifacts remains challenging. Human evaluation, while effective, is very costly and time-consuming. Traditional automated metrics like BLEU rely on high-quality references and struggle to capture nuanced aspects of software quality, such as readability and usefulness. In response, the LLM-as-a-Judge paradigm, which employs LLMs for automated evaluation, has emerged. Given that LLMs possess strong coding abilities and reasoning skills, they hold promise as cost-effective and scalable surrogates for human evaluators. Nevertheless, \toolname research in the SE community is still in its early stages, with many breakthroughs needed. 

This forward-looking SE 2030 paper aims to steer the research community toward advancing \toolname for evaluating LLM-generated software artifacts, while also sharing potential research paths to achieve this goal. We provide a literature review of existing SE studies on \toolname and envision these frameworks as reliable, robust, and scalable human surrogates capable of evaluating software artifacts with consistent, multi-faceted assessments by 2030 and beyond. To validate this vision, we analyze the limitations of current studies, identify key research gaps, and outline a detailed roadmap to guide future developments of \toolname in software engineering. While not intended to be a definitive guide, our work aims to foster further research and adoption of \toolname frameworks within the SE community, ultimately improving the effectiveness and scalability of software artifact evaluation methods.

  \end{abstract}



\keywords{Large Language Models, Software Engineering, LLM-as-a-Judge, Research Roadmap}

\received{20 February 2025}
\received[revised]{}
\received[accepted]{}

\maketitle

\section{Introduction}

With significant advances in Artificial Intelligence (AI)—particularly in Deep Learning (DL)~\cite{lecun2015deep,goodfellow2016deep} and Large Language Models (LLMs)~\cite{OpenAI_GPT4, deepseek2405deepseek, anthropic2023claude}—various fields have increasingly leveraged LLMs to develop automated solutions, notably within Software Engineering (SE)~\cite{10.1145/3695988,10.1145/3712005,fan2023large,wang2024software,10.1145/3708525}. Demonstrating exceptional performance across a spectrum of longstanding challenges, ranging from code generation~\cite{zhuo2024bigcodebench, liu2024exploring} and summarization~\cite{ahmed2024automatic, sun2024source} to program translation~\cite{yuan2024transagent, pan2023understanding} and repair~\cite{10.1109/ICSE48619.2023.00125,jin2023inferfix}, LLMs have powered a series of new applications and tools that streamline the software development process, such as GitHub Copilot~\cite{githubGitHubCopilot}, Cursor Code Editor~\cite{cursor}, and more.

However, these useful innovations have also brought new challenges in SE research, among which a key question is:
\begin{tcolorbox}[width=\textwidth, box align=center, left=1.2cm, top=0pt, bottom=0pt, right=0.5cm, colback=white, frame empty]
{\it How can we comprehensively and scalably evaluate the quality of LLM-generated software artifacts?}
\end{tcolorbox}

Assessing these LLM-generated software artifacts through experienced developers, i.e., {\it human evaluation}, is an ideal approach, as software quality should be judged based on its effectiveness in real-world scenarios~\cite{tian2005software}. However, rigorous human evaluation poses significant challenges.
A survey of SE researchers found that 84\% agree that human evaluation is problematic, primarily due to time constraints, the need for specialized practical knowledge, and the high cost involved~\cite{buse2011benefits}. Additionally, human evaluators may experience fatigue and reduced focus, further compromising the quality of the evaluation and delaying the evaluation process~\cite{kumar2024llms}.

Therefore, researchers have increasingly adopted automated metrics for evaluating LLM-generated software artifacts~\cite{hu2022correlating,papineni2002bleu, lin2004rouge}, given their relatively low cost and scalability. However, traditional automated metrics come with their own limitations. For example, Pass@$k$, a widely used metric for measuring code correctness, involves executing the first $k$ generated code snippets against unit tests. Although useful, it still demands significant human effort to design comprehensive test suites and manually configure execution environments. Similarly, the effectiveness of text similarity-based metrics like BLEU~\cite{papineni2002bleu} and CodeBLEU~\cite{ren2020codebleu}, along with embedding-based metrics like BERTScore~\cite{zhang2019bertscore} and CodeBERTScore~\cite{zhou2023codebertscore}, rely heavily on high-quality references that are typically annotated by human experts.
Moreover, these methods not only fail to capture nuanced aspects of software artifacts—such as naturalness, usefulness, and adherence to best practices—as humans would, but they also often lead to misalignment with human judgment, as evidenced in several studies~\cite{hu2022correlating, kumar2024llms, roy2021reassessing, evtikhiev2023out}.
Consequently, achieving comprehensive evaluation while ensuring scalable automation remains a persistent challenge in assessing the quality of LLM-generated software artifacts.

Recently, the remarkable success of LLMs has inspired the emergence of the LLM-as-a-Judge (LLM-J) paradigm~\cite{zheng2024judging}, which extends LLMs' role from content generation to comprehensive content evaluation, positioning them as scalable and cost-effective surrogates for human evaluators. This paradigm is driven by several key attributes of LLMs. First, numerous studies have shown that LLMs exhibit both impressive coding abilities~\cite{liu2024deepseek, lozhkov2024starcoder, li2023starcoder} and human-like reasoning skills~\cite{patil2025advancing, ivanova2025evaluate,guo2025deepseek}. 
Additionally, since LLMs are often trained through reinforcement learning from human feedback (RLHF)~\cite{bai2022training}, their outputs closely align with human expert judgments~\cite{githubcopilot2025, chen2021evaluating}. Moreover, unlike human evaluators, LLMs do not experience fatigue or reduced efficiency from prolonged work, allowing them to maintain consistent performance over extended periods. These attributes collectively make LLMs highly suitable in content evaluation tasks.

In this paper, we argue that \toolname systems offer a promising solution to address the limitations of both costly human evaluation and traditional automated metrics in SE. Our research community is beginning to recognize the potential of LLM-as-a-Judge, as evidenced by a few empirical studies and emerging methodologies on this topic such as~\cite{wang2025can,wu2024can,ahmed2024can}. Nevertheless, these studies represent only initial progress, and the field remains in its early stages. Many breakthroughs are still needed, and the overall landscape is far from being fully mapped, especially as we look to 2030 and beyond. This necessitates this paper to redirect the research community’s focus by presenting a vision for the future of \toolname and a roadmap that inspires further progress.

In the following sections, this paper first provides a formal definition of the \toolname concept (\autoref{sec:background}). We then review key historical milestones and recent advances in \toolname for software engineering (\autoref{sec:review}). Looking beyond 2030, we envision \toolname systems as reliable, robust, and scalable human surrogates capable of evaluating a wide range of software artifacts while providing consistent, multi-faceted assessments.
To validate this vision, we identify key limitations and research gaps, outlining a roadmap with specific research directions and potential solutions for the SE research community to pursue (\autoref{sec:roadmap}). In summary, our study makes the following key contributions:
\begin{itemize}[leftmargin=*]
\item We conduct a comprehensive review of 16 primary studies on \toolname in software engineering.
\item We analyze the limitations of existing research on \toolname in SE, identifying major challenges and research gaps.
\item We present a forward-looking vision for \toolname in SE by 2030 and beyond, proposing a detailed research roadmap with key opportunities for future exploration.
\end{itemize}
\section{Definition of LLM-as-a-Judge}
\label{sec:background}

As a recent emerging topic, various definitions of \toolname are currently proposed and not yet unified. 
This paper was inspired by the definition proposed by Li et al.~\cite{li2024llms}. 
Informally speaking, \toolname utilizes LLMs as evaluative tools to assess software artifact based on predefined evaluation criteria.
Formally speaking, \toolname is defined as follows:

\begin{equation}
 E(\mathcal{T}, \mathcal{C}, \mathcal{X}, \mathcal{R}) \rightarrow (\mathcal{Y}, \mathcal{E}, \mathcal{F})
\end{equation}

The function \( E \) is the core evaluation mechanism powered by LLMs. It takes the following key \textbf{inputs}:

\begin{itemize}[leftmargin=*]
    \item \( \mathcal{T} \) – Evaluation Type. It mainly covers three types: \textit{point-wise}, \textit{pairwise}, and \textit{list-wise}.
    \begin{itemize}
        \item \textbf{Point-wise Evaluation:} Function \( E \) evaluates a single candidate content independently and often provides a score or categorical classification.
        \item \textbf{Pair-wise Evaluation:} Function \( E \) compares two candidates and determines their relative quality.
        \item \textbf{List-wise Evaluation:} Function \( E \) ranks multiple (more than 2) candidates.
    \end{itemize}
    \item \( \mathcal{C} \) – Evaluation Criteria. It describes the assessment aspects, e.g., correctness, helpfulness, readability, etc.
    \item \( \mathcal{X} \) – Evaluation Item. It is the content to be judged. Note that \( \mathcal{X} \) can be a single candidate or a set of candidates \( (x_1, x_2, ..., x_n) \), where \( x_i \) is the \( i^{th} \) candidate to be evaluated.
    \item \( \mathcal{R} \) – Optional reference information. For instance, a code implementation that successfully fulfills the requirements of a coding task.

\end{itemize}

Given these inputs, \( E \) can produce three primary \textbf{outputs}:

\begin{itemize}[leftmargin=*]
    \item \( \mathcal{Y} \) – Evaluation Result. \( \mathcal{Y} \) can take one of the following forms:
    \begin{itemize}[leftmargin=*]
        \item \textbf{Graded Evaluation:} Each candidate is assigned a score, either numerical (discrete or continuous) or categorical.      
        \item \textbf{Rank Ordering:} Candidates  \( (x_1, x_2, ..., x_n) \) are ranked based on their assessed quality.
        \item \textbf{Best-Choice Selection:} Function \( E \) selects the most suitable candidate ($x_i$) from the given set \( (x_1, x_2, ..., x_n) \).
    \end{itemize}

    \item \( \mathcal{E} \) – Evaluation Explanation. Optionally provides reasoning and justification for the evaluation.  
    \item \( \mathcal{F} \) – Feedback. Optionally offers constructive suggestions for improving the evaluated item \( \mathcal{X} \).  
\end{itemize}

\textbf{Difference to previous work. } Note that this definition of \toolname is stricter than the definition proposed by Wang et al.~\cite{wang2025can}. Wang et al. broadly define \toolname{} as any method that leverages any features of LLM to evaluate content quality, which includes embedding-based approaches, such as BERTScore~\cite{zhang2019bertscore} and CodeBERTScore~\cite{zhou2023codebertscore}. 
Embedding-based methods evaluate content similarity by converting both the evaluated content and references into numerical representations (LLM embeddings).
However, we argue that their definition is better characterized as \textit{LLM-based evaluation} rather than \textit{\toolname{}}. The key distinction is that we envision \toolname{} as a true surrogate for human evaluation. In our view, an \toolname system should be capable of independently assessing content without requiring a strict reference. 
Moreover, it should be able to perform a wide spectrum of nuanced assessments (e.g., usefulness, adequacy, and conciseness) and can ideally provide justifications or constructive feedback. 
In contrast, embedding-based methods rely solely on the numerical representation of assessing content similarity and inherently lack these capabilities. Consequently, we do not include them in our definition and instead leave a stricter definition to previous work.

\section{Literature Review}
\label{sec:review}

This section, we provide an overview of studies that evaluate the effectiveness of LLM-as-a-Judge in SE tasks, with Table~\ref{tab:se-tasks} providing a structured summary that maps each study to its corresponding SE task. Following this, we present a detailed discussion of the evaluation objectives of these studies.

\begin{table}[]
   \centering
   \caption{Literature Review on the Applications of LLM-as-a-Judge in Software Engineering}
   \label{tab:se-tasks}
   \begin{tabular}{@{}ll@{}}
   \toprule
   Task                              & Reference                                                                                                                                                                                  \\ \midrule
   Code Generation                   & \cite{wang2025can, tong-zhang-2024-codejudge, ahmed2024can, patel2024aime, tan2024judgebench, zhao-etal-2025-codejudge, zhuo2023ice, sollenberger2024llm4vv, weyssow2024codeultrafeedback, xu2024human} \\
   Code Summarization                & \cite{wang2025can, ahmed2024can, farchi2024automatic, wu2024can, weyssow2024codeultrafeedback}                                                                                             \\
   Bug Report Summarization          & \cite{kumar2024llms}                                                                                                                                                                       \\
   Code Translation                  & \cite{wang2025can}                                                                                                                                                                         \\
   Question Answering                & \cite{zheng2024judging}                                                                                                                                                                    \\
   Requirements Causality Extraction & \cite{ahmed2024can}                                                                                                                                                                        \\
   Code Patches Generation           & \cite{yadavally2025large, ahmed2024can, li2024cleanvul}                                                                                                                                    \\ \bottomrule
   \end{tabular}
   \end{table}

\subsection{Code Generation}
Effectively and automatically evaluating the quality of generated code is one of the most critical problems in software engineering.
Before the introduction of \toolname, automated code evaluation primarily relied on test-based method, i.e., executing the code based on pre-defined test cases~\cite{zhuo2024bigcodebench, chen2021evaluating, hendrycks2021measuring}. 
However, passing all the test cases does not necessarily mean the quality of the code is good, as the test cases may not comprehensively cover all the edge cases~\cite{dou2024s}.
Constructing test cases can also be particularly challenging for certain tasks~\cite{tong-zhang-2024-codejudge}, such as AI model training, web scraping, etc. 
When test cases are unavailable, many studies rely on reference-based metrics~\cite{dehaerne2022code, zheng2023codegeex}, such as CodeBLEU~\cite{ren2020codebleu} and CodeBERTScore~\cite{zhou2023codebertscore}. However, the quality of the reference code restricts the effectiveness of reference-based metrics. Prior research~\cite{evtikhiev2023out} highlighted that reference-based metrics frequently misalign human judgments in code generation. 

Consequently, a substantial body of research has explored the effectiveness of LLMs in evaluating code generation~\cite{wang2025can, zheng2024judging, tong-zhang-2024-codejudge, patel2024aime, tan2024judgebench, zhao-etal-2025-codejudge, zhuo2023ice}. We summarize three main characteristics of \toolname that are different from the test-based and reference-based metrics:

\begin{itemize}[leftmargin=*]
    \item \textbf{Execution-free:} It does not require composary execution of the code.
    \item \textbf{Reference-free:} Unlike reference-based metrics, \toolname does not require a reference code. Even though we can still optionally provide the reference code. However, as demonstrated by Zhuo et al.~\cite{zhuo2023ice}, providing the reference does not generally improve the code evaluation performance.
    \item \textbf{Multi-Facet Evaluation:} \toolname assesses intrinsic code qualities that traditionally required human judgment, such as readability and usefulness.
\end{itemize}

Among existing studies, Zhuo et al.~\cite{zhuo2023ice} and Weyssow et al.~\cite{weyssow2024codeultrafeedback} utilize GPT-3.5~\cite{openai2023gpt35}, and Xu et al.~\cite{xu2024human} use GPT-4 to annotate code snippets. 
CodeJudge~\cite{tong-zhang-2024-codejudge} first leverages a detailed taxonomy of common programming errors to guide the LLM in analyzing the generated code.
Then, CodeJudge summarizes the analysis report to produce its final judgment.
Wang et al.~\cite{wang2025can} experiment with a series of general-purpose \toolname methods from the natural language processing (NLP) domain in code generation evaluation, including BatchEval~\cite{yuan2023batcheval}, GPTScore~\cite{fu2023gptscore}, and G-Eval~\cite{liu2023g}.
Ahmed et al.~\cite{ahmed2024can} investigate LLM's ability to evaluate variable name-value inconsistency and function similarities. 
Gu et al.~\cite{gu-etal-2024-counterfeit} focus on analyzing LLM's judgments on counterfeit code samples, i.e., incorrect programs generated by language models that pass weak but non-trivial correctness checks.
Zhao et al.~\cite{zhao-etal-2025-codejudge} empirically evaluate the performance of 12 LLMs on code generation evaluation. 
Patel et al.~\cite{patel2024aime} hypothesis that a combination of multiple evaluators can approximate the optimal evaluation. Thus, they propose the AIME framework, which employs multiple LLMs to evaluate different aspects of the code, including code correctness, readability, runtime performance, etc. 

Further, we summarize the common evaluation criteria used in the literature for code generation. We categorize them into two main criteria: \textit{Code Functionality} and \textit{Code Quality}. 

\noindent\textbf{Code Functionality:} This criterion assesses the overall effectiveness, maintainability, and clarity of the code beyond its functional correctness. It evaluates intrinsic code attributes. Key aspects include:

\begin{itemize}[leftmargin=*, label=-]
   \item \textbf{Execution Stability:} \toolname's evaluation process does not mandate code execution, meaning that running the code is not a required step.
   \item \textbf{Functional Correctness:} Verifies whether the code produces the expected output according to the task description.
   \item \textbf{Fault Tolerance \& Error Handling:} Evaluates how the code manages edge cases, exceptions, invalid inputs, and unexpected failures.
\end{itemize}
  
\noindent \textbf{Code Quality:} These criteria evaluate intrinsic attributes such as readability and adherence to best practices.  
\begin{itemize}[leftmargin=*, label=-]
    \item \textbf{Complexity \& Efficiency:} Analyzes computational complexity, execution time, and memory usage.
    \item \textbf{Helpfulness:} Evaluates whether the code contributes meaningfully to solving the problem. This includes distinguishing partially correct solutions from completely unuseful and irrelevant code.
    \item \textbf{Readability:} Measures how easily the code can be understood, including aspects like fluency, clarity, and conciseness.
    \item \textbf{Stylistic Consistency:} Ensures adherence to established coding standards, formatting guidelines, and best practices.
    \item \textbf{Reference Similarity:} Assesses how closely the generated code aligns with a given reference implementation.
    \item \textbf{Minor Anomalies \& Warnings:} Identifies non-critical issues such as redundant code or unused variables that do not affect execution.
\end{itemize}

\subsection{Code Change}  
Beyond code generation, LLM-as-a-Judge has been explored for evaluating code modifications.  
Ahmed et al.~\cite{ahmed2024can} use \toolname to determine whether a code patch effectively resolves a static analysis warning.  
Li et al.~\cite{li2024cleanvul} apply LLMs to assess whether a patch successfully fixes a vulnerability, while Yadavally et al.~\cite{yadavally2025large} leverage LLMs to predict test outcomes of code patches.

\subsection{Software Documentation Summarization}
LLM-based judges have also been applied to assess natural language software documentation.
Ahmed et al.~\cite{ahmed2024can} and Wu et al.~\cite{wu2024can} introduce LLM judges to evaluate code summaries.
Wu et al.~\cite{wu2024can} introduce CODERPE, a framework that utilizes multiple LLMs and assigns LLMs with different roles to evaluate code summaries from various perspectives. These roles include a code reviewer, the original code author, a code editor, and a system analyst. 
In addition, Kumar et al.~\cite{kumar2024llms} utilized LLMs to evaluate the quality of bug report titles and summaries.
For code summarization, the common evaluation criteria include:
\begin{itemize}[leftmargin=*]
    \item \textbf{Language-related:} Ensure clear, well-structured sentences with precise and appropriate wording.
   \item \textbf{Content-related:} Summarize the core functionality and logic concisely, providing sufficient detail without including unnecessary information.
   \item \textbf{Effectiveness-related:} Evaluate whether the summary is useful for developers and enhances their understanding of the code.
\end{itemize}

\subsection{Other SE Tasks}  

LLM-as-a-Judge has also been explored in other software engineering tasks.  
Zheng et al.~\cite{zheng2024judging} apply this approach to assess the quality of programming-related question answering.  
Wang et al.~\cite{wang2025can} investigate its effectiveness in assessing code translation quality. In addition, Ahmed et al.~\cite{ahmed2024can} leverage LLMs to evaluate causal relationships extracted from natural language requirements.

\section{The Road Ahead}
\label{sec:roadmap}

This section outlines a research roadmap for achieving scalable, reliable, and effective LLM-as-a-Judge systems in software engineering. 
We begin by outlining the key limitations in current approaches and then propose concrete research opportunities and actionable steps that can significantly broaden the applicability of \toolname{} across diverse SE contexts, steering us toward our envisioned future.

\subsection{More Empirical Evaluation and Benchmarking}

\textbf{Limitation 1: Lack of High-Quality and Large-Scale Human-Annotated Benchmarks. }
In Section \ref{sec:review}, we reviewed numerous studies evaluating the effectiveness of LLM-as-a-Judge in software engineering. Benchmarks such as HumanEval~\cite{chen2021evaluating}, HumanEval-XL~\cite{peng2024humaneval}, and CoNaLa~\cite{yin2018learning} provide useful test cases to evaluate the ability of LLMs in assessing code correctness. However, these benchmarks fall short when assessing more nuanced aspects like helpfulness, readability, and alignment with human judgment.
A critical limitation in existing \toolname of SE is their reliance on small-scale datasets to measure human alignment~\cite{wang2025can, wu2024can}, often comprising only a few hundred samples.
For example, Wang et al.~\cite{wang2025can} conducted experiments on three SE tasks, i.e., Code Translation, Code Generation, and Code Summarization, using a total of only 450 samples. Similarly, Ahmed et al.~\cite{ahmed2024can} evaluated code summarization using just 420 samples. While these sample sizes may be sufficient for demonstrating statistical significance, larger-scale benchmarks are also essential to ensure generalizability and mitigate threats to external validity.

\vspace{0.1cm}
\noindent \textbf{Limitation 2: Inconsistent Empirical Findings. }Related to these small-scale experiments, another major challenge is the inconsistency in empirical findings.
A few works reached varying conclusions due to discrepancies in dataset selection, prompting strategies, and LLM configurations.
For instance, Wang et al.~\cite{wang2025can} found that traditional evaluation metrics (e.g., ROUGE~\cite{lin2004rouge}, METEOR~\cite{banerjee2005meteor}, and ChrF++~\cite{popovic2017chrf++}) significantly outperformed LLM-as-a-Judge methods when aligning with human judgment for code summarization. In contrast, Wu et al.~\cite{wu2024can} reported that the \toolname method surpassed conventional metrics in code summarization. These conflicting results highlight the need for a standardized, large-scale empirical study to enable fair and meaningful comparisons across studies.

\vspace{0.1cm}
\noindent \textbf{Limitation 3: Lack of Empirical Findings in Bias of \toolname systems in SE.} Despite the growing adoption of \toolname in SE, software artifacts and data may contain biases~\cite{9825855,9609175}. LLMs are already known to be susceptible to various biases and fairness issues. For example, Position bias~\cite{li-etal-2024-split}, where LLMs may favor responses based on their positioning within the text when making comparison; Verbosity bias~\cite{jiao2024enhancing}, where LLMs show a tendency for longer, verbose responses, regardless of their qualities; Egocentric Bias~\cite{ye2024justice}, where LLMs may overrate AI-generated solutions compared to human-written code. However, currently, there is a lack of thorough empirical investigation into the bias of \toolname systems in SE.

\vspace{0.1cm}
\noindent\textbf{\textit{Opportunity: Develop Comprehensive and Diverse Benchmarks.}} Future research should work on creating large-scale, multi-dimensional benchmarks that capture the complexity of real-world software engineering tasks. Key steps include:
\begin{enumerate}[leftmargin=*]
    \item \textit{Expert Annotations and Quality Control:} Engage a diverse pool of expert programmers to provide high-quality annotations. This annotation process should focus more on the nuanced criteria beyond code correctness, such as readability, helpfulness, and maintainability. In the beginning, define clear annotation guidelines that describe these nuanced dimensions to eliminate ambiguity.
    During the process, cross-validation and consensus mechanisms should be conducted to ensure the reliability of the annotation.

    \item \textit{Dataset Expansion:} Expand existing datasets to include a larger volume of high-quality expert-annotated samples. These benchmarks should encompass SE tasks at varying difficulty levels, various programming topics, and multiple programming languages. Additionally, new benchmarks should be extended to evaluate a broader spectrum of SE artifacts. For instance, expert evaluations could be gathered for software requirements documentation, system design, and API documentation.
    In the long term, we may also consider updating human preferences in the benchmarks to reflect evolving SE practices.
\end{enumerate}

\vspace{0.1cm}
\noindent\textbf{\textit{Opportunity: Comprehensive Empirical Evaluation in SE Tasks.}} Ultimately, these benchmarks will lay a solid foundation for understanding the strengths and weaknesses of LLM-as-a-Judge systems in SE. Large-scale experiments that systematically compare LLM-as-a-Judge methods with traditional metrics, while the impacts of design choices, such as prompting strategies and LLM configuration, will be investigated. Further, a thorough investigation into the bias and fairness of \toolname systems in SE will be conducted. 

\subsection{Better Judgment From Internal Factors}

\noindent \textbf{Limitation 4: Inadequate SE Domain-Specific Expertise in LLMs.} While ideally, human evaluations should be conducted by experienced developers. Like human evaluators, LLM must exhibit a deep understanding of the task and the relevant domain knowledge to provide an accurate evaluation.
Recent research highlights that LLMs have remarkable coding abilities, which builds confidence in their ability to assess tasks like code summarization and generation.
However, research also indicates that some complex coding tasks are still challenging for LLMs~\cite{zhuo2024bigcodebench, zhao2024commit0}. For instance, Zhao et al.~\cite{zhao2024commit0} observed that current LLMs cannot generate a complete library from scratch, which also implies LLMs' ability to evaluate library quality is still limited. Similar limitations extend to other critical areas in software engineering, such as software design, formal verification, distributed systems debugging, etc.
Beyond gaps in SE-specific knowledge, LLMs may also struggle with discerning subtle differences among alternative solutions. 
As Zhao et al.~\cite{zhao-etal-2025-codejudge} highlighted, an LLM's ability to generate correct code does not guarantee that it can effectively evaluate alternative implementations for the same task.

\vspace{0.1cm}
\noindent\textbf{\textit{Opportunity: Enhancing SE Expertise in LLM.}} Future research can strengthen the SE domain-specific expertise of LLMs. A key approach is training LLMs on richer and more high-quality SE datasets~\cite{liu2024datasets}. Enhancing SE expertise in LLMs offers several benefits. For instance, LLMs with stronger formal verification capabilities can more accurately assess code correctness. Further, research can expand \toolname's applicability to a broader range of SE artifacts, such as Docker files, UML diagrams, and formal specifications. 

\vspace{0.1cm}
\noindent\textbf{\textit{Opportunity: Embedding Expert Tacit Knowledge.}}
Another critical research direction for enhancing judgment involves capturing the tacit and procedural knowledge that human experts develop over years of experience, i.e., their intuition and unspoken decision-making processes. We can employ methods such as structured interviews~\cite{segal2006structured}, think-aloud protocols~\cite{zhang2019think}, and cognitive task analyses~\cite{schraagen2000cognitive}, where experts are asked to articulate their reasoning while conducting real-world evaluations. This approach enables the systematic extraction of procedural knowledge that is typically not documented in software artifacts. Once captured, this knowledge can be integrated into LLM training through techniques like reinforcement learning~\cite{bai2022training} or neuro-symbolic methods~\cite{kwon2023neuro}. 

\subsection{Better Judgment Through External Factors}

\noindent \textbf{Limitation 5: Reliance on Internal Evaluation Mechanisms.} Current works tend to depend solely on LLMs' own abilities for evaluation. This shortfall underscores the need for additional mechanisms to assist \toolname in decision-making.

\vspace{0.1cm}
\noindent\textbf{\textit{Opportunity: Integrating SE tools with \toolname.}}
As humans rely on other tools to assist their judgment, \toolname can incorporate outputs from various analysis tools. Integrating tools such as static analyzers, formal verification frameworks, and model checkers into the evaluation pipeline—or conversely, embedding \toolname into other tools like Integrated Development Environments (IDEs)~\cite{10.1145/3708525,10.1145/3639475.3640097,10.1145/3551349.3556964}—can add additional layers of validation.

\vspace{0.1cm}
\noindent\textbf{\textit{Opportunity: Human-in-the-loop.}}
For tasks that cannot be reliably automated by LLMs, we need to develop collaborative and interactive methods to include human oversight. Also, identifying the optimal ratio of LLM evaluators to human experts remains another challenge. Moreover, enhancing LLMs with the ability to assess their own confidence levels is crucial. When LLM generates low-confidence ratings, it should automatically flag these cases for human review.

\subsection{Addressing The Security Issues of Using \toolname}

\noindent \textbf{Limitation 6: Insufficient Research on Adversarial Threats and Defensive Methods in SE.}
Adversarial attacks~\cite{yang2022natural, 9825775, yefet2020adversarial} on LLM-as-a-Judge systems pose significant risks by subtly manipulating software artifacts to alter evaluation outcomes. For example, attackers may obfuscate code by injecting misleading comments or rearranging segments to mask critical vulnerabilities. Adversaries might also modify bug reports or commit messages with deceptive context to influence LLM's code quality, maintainability, and security assessment. However, at the current stage, we notice that this threat for \toolname is under-explored in the SE community.

\vspace{0.1cm}
\noindent\textbf{\textit{Opportunity: Adversarial Testing and Robust Defensive Mechanisms.}} Future research should adopt approaches that develop both adversarial methods and defensive strategies. On the one hand, developing adversarial training techniques~\cite{gong2022curiosity,xhonneux2024efficientadversarialtrainingllms} is crucial to enhance the LLM's robustness, where crafted adversarial examples are incorporated into the training process, 
On the other hand, additional robust defensive mechanisms, such as anomaly detection~\cite{song-etal-2025-confront,10431665}, should be integrated to detect and mitigate the threat of adversarial attacks proactively.  

\vspace{-0.15cm}

\section{Conclusion}

This forward-looking SE 2030 paper first present a brief overview of the current landscape of LLM-as-a-Judge systems, identifying their limitations while highlighting the associated opportunities and presenting a future vision for their adoption in the SE community. To realize this vision by 2030 and beyond, we propose a research roadmap outlining the specific paths and potential solutions that the research community can pursue. Through these efforts, we aim to encourage broader participation in the LLM-as-a-Judge research journey.

\bibliographystyle{ACM-Reference-Format}
\bibliography{reference}

\end{document}